\documentclass[preprint,review,12pt]{elsarticle}

\usepackage{ar}
\usepackage{lineno,hyperref}
\usepackage{soul}
\usepackage[dvipsnames]{xcolor}
\modulolinenumbers[5]
\usepackage{epsfig,graphicx}
\usepackage{amssymb}
\usepackage{amsmath}
\usepackage{comment}
\usepackage{enumerate}
\usepackage{listings}
\usepackage{tabularx}
\usepackage{booktabs}
\usepackage{subcaption}
\usepackage[T1]{fontenc}

\biboptions{numbers,sort&compress}

\journal{Arxiv}

\begin{document}

\begin{frontmatter}

\title{Digital Models of Widely-Used Powder Characterisation Tools and their Application in DEM Calibration \\ Part III: Packing and Flow Testers}



\author[add1]{B.D. Jenkins\corref{cor}}
\author[add1]{A. L. Nicu\c{s}an}
\author[add1]{D. Werner}
\author[add1]{J.P.K. Seville}
\author[add2]{A. Neveu}
\author[add3]{G. Lumay}
\author[add2]{F. Francqui}
\author[add1]{C.R.K. Windows-Yule}
\cortext[cor]{bdj746@student.bham.ac.uk}

\address[add1]{School of Chemical Engineering, the University of Birmingham, Edgbaston, Birmingham, B15 2TT, UK}
\address[add2]{Granutools SPRL, Rue Jean Lambert Defr$\mathrm{\hat{e}}$ne 107, 4340 Awans, Belgium}
\address[add3]{Grasp laboratory, CESAM research unit, University of Liège, Place du 20 Août 7, 4000 Liège, Belgium}

\begin{abstract}
Accurate calibration remains a critical challenge in the field of discrete element method (DEM) simulations. To work towards more standardised and rigorous methods, this research contributes to a growing open-source library of digital models that faithfully replicate powder characterisation instruments. This paper expands that library by developing and validating digital models for two widely used instruments: the GranuFlow flow through an orifice tester and the GranuPack tapped density analyser.
These models are designed to replicate the experimental instruments while incorporating computational optimisations to improve efficiency without compromising on accuracy. An example application of the digital models is shown through a calibration case study for two example materials: micro-crystalline cellulose (MCC) and ion exchange resin.
The calibrated models show strong agreement with experimental data through cross-validation, yielding a relative percentage error of 0.57\% for the Hausner ratio (the ion exchange resin) and 4.17\% for a Beverloo fitting parameter (MCC). The addition of these validated, open-source models provides the DEM community with further models for more robust, multi-instrument calibration, and potential for more insight into bulk behaviours under different conditions.
\end{abstract}

\end{frontmatter}

\newpage

\section{Introduction}
\label{sec_intro}

In previous instalments of this series, a range of digital models that faithfully replicate widely-used powder characterisation instruments was developed, including shear cells and angle of repose testers \cite{Werner2025Open-sourceRheology, WernerDigitalTesters}. These initial models allow for the analysis of powder behaviour under consolidated, unconfined static, and dynamic regimes. However, they do not investigate granular dynamics such as particle rearrangement, compaction or arch formations. 

To address these gaps and to expand the range of open-source digital models, this work develops digital models of a tapped density analyser, GranuPack (Granutools, Belgium), and a flow through an orifice tester, GranuFlow (Granutools, Belgium). Expanding this library of digital twins is critical for developing a more comprehensive understanding of a material by simulating material behaviour across a wider range of conditions. Furthermore, employing multiple digital instruments provides a more robust framework for the calibration of the discrete element method (DEM). Since each characterisation test is uniquely sensitive to different inter-particle contact parameters (such as friction and cohesion) combining their results provides stronger constraints on the parameter space \cite{Jenkins2024TheProperties}. This multi-instrument approach is essential for identifying a single, unified set of DEM parameters that accurately represents the material across diverse applications \cite{Elskamp2017AParticles}.

The flow through an orifice tester and tapped density analyser were chosen to expand the suite of digital models as they measure powder behaviour in a fundamentally distinct way from those assessed by shear cells or angle of repose testers. The orifice flow test characterises gravity-driven flow under low-stress conditions, providing critical insights into the unconfined flowability of a bulk material and its tendency to form arches. The tapped density analyser quantifies powder consolidation/compaction and packing behaviour due to an energy input, which helps to understand particle rearrangement behaviour in a bulk material. By simulating instruments that investigate these different mechanics, the new models provide a unique insight into bulk materials compared to the previous digital models developed and facilitate the meaningful calibration of simulations of a still wider range of particulate systems, such as vibropacking processes or hopper discharge. 

With the addition of the digital models developed in this paper, an almost full range of particle behaviours can be investigated across this library of DEM simulations.

\section{Materials and Methods}

    \subsection{Experimental Materials}
    \label{sec:experimental_materials}

    To demonstrate the digital models in a calibration example use case in Section \ref{sec:granuflow_calibration} and Section \ref{sec:granupack_calibration}, two materials were measured in the experiments: ion exchange resin, and micro-crystalline cellulose (MCC) granules.

    The MCC spheres (Vivapur 1000) had a median diameter ($d_{50}$) of 1.2~±~0.1~mm and mean sphericity ($\phi$) of 0.940~±~0.004; and the ion exchange resin (Purolite PCA433), with a measured median diameter of 290~±~70~$\mu$m and sphericity of 0.944~±~0.006. The material densities were 1460~kgm$^{-3}$ and 960~kgm$^{-3}$, respectively \cite{Werner2025Open-sourceRheology, WernerDigitalTesters}. Angular sand, a material used in previous papers in this series \cite{Werner2025Open-sourceRheology, WernerDigitalTesters}, was excluded from the current study as the large particle size blocked flow in the flow through an orifice instrument, an issue observed both in physical experiments and in the corresponding simulations. This meant a steady flow could not be obtained and thus a proper measurement could not be made. 
    
    Following the standard operating procedures for each instrument, the flow rate through a standardised set of orifice sizes was measured using the GranuFlow to calculate the Beverloo fitting parameter `$C$', and the tapped density curves were acquired with the GranuPack to calculate the Hausner ratio ($H_r$). The details of calculating these quantities will be given in the background section of each instrument (Section \ref{sec:real_granuflow} and \ref{sec:tapped_density}).
    
    The two materials were measured in both instruments to have a measurement to use for calibration and later a measurement to validate (in Section \ref{sec:cross_validation}), these experimental measurements are summarised in Table \ref{table:experimental_measurements}.

       \begin{table}[ht!]
    \def\arraystretch{1.5}
    \centering
    \caption{Summary of experimental measurement results.}
    \label{table:experimental_measurements}
        \begin{tabularx}{\linewidth}{ | >{\raggedright\arraybackslash}X || >{\centering\arraybackslash}X | >{\centering\arraybackslash}X | }
        \hline
        \textbf{Bulk Measurement} & \textbf{MCC} & \textbf{Ion Exchange Resin} \\
        \midrule\midrule
        Beverloo Fitting Term ($C$) & 0.0575 & 0.0586 \\
        \hline
        Hausner Ratio ($H_r$) & 1.062 & 1.052 \\
        \hline
        \end{tabularx}
    \end{table}


    \subsection{Simulation Details}
    \label{sec:sim_details}

    After discussing the background of the experiment and details of the digital models in Section \ref{sec:digital_models} this paper will conduct an example use of the digital models for a simple calibration of DEM contact parameters. The general details of the simulations are presented here.
    
    All simulations in this study were performed with the PICI-LIGGGHTS \cite{UniversityofBirminghamPositronImagingCentre2022UoBPICI-LIGGGHTS-3.8.1} open-source package, which is an improved version of the LIGGGHTS DEM software \cite{Kloss2012ModelsCFD-DEM}. This distribution is publicly available on this GitHub repository: \href{https://github.com/uob-positron-imaging-centre/PICI-LIGGGHTS}{https://github.com/uob-positron-imaging-centre/PICI-LIGGGHTS}. While this paper focuses on PICI-LIGGGHTS, the digital models library now includes digital models of widely-used powder characterisation instruments across multiple software including BlazeDEM \cite{Govender2016Blaze-DEMGPU:Architecture}, Rocky DEM (Ansys) and EDEM (Altair). All of these digital models are available to freely use and are downloadable from the Granutools Gitlab (\href{https://gitlab.com/gdigitaltwins}{https://gitlab.com/gdigitaltwins}). The PICI-LIGGGHTS versions are also available from the University of Birmingham Positron Imaging Centre GitHub repository (\href{https://github.com/uob-positron-imaging-centre}{\color{black} https://github.com/uob-positron-imaging-centre}).

    To model the mechanics of particle contacts the Hertz-Mindlin model \cite{DiRenzo2004ComparisonCodes} is used to implement elastic collisions and energy loss via the coefficient of restitution ($\varepsilon$) and sliding friction ($\mu_s$) via the Coulomb limit as well as the Constant Directional Torque (CDT) model \cite{Ai2011AssessmentSimulations} for rolling friction ($\mu_r$) in the PICI-LIGGGHTS simulations. It is important to note that the digital models presented here are flexible, allowing for the integration of alternative force models as required.

    The input parameters for MCC and ion exchange resin for the simulations are listed in Table \ref{table:sim_params}. Material properties, including particle diameter and density, were based on direct experimental measurements. The median particle diameter ($d_{50}$) of MCC was kept the same in the simulations as the experimentally measured value, however, the ion exchange resin was coarse grained by a factor of two so; $d_{50}=0.58$mm. This coarse graining was done to reduce the computational expense. Coarse graining particles in DEM up to 20 times has been shown to provide quantitatively accurate results to experiments in positron emission particle tracking (PEPT) experiments \cite{Che2023EvaluationMeasurements}. The coefficient of restitution was determined empirically via simple impact tests \cite{Lorenz1997MeasurementsParticles,Muller2013ComplexBall, Inoue1996GrindingMethod}.

    A Poisson's ratio of 0.3 was assumed for both materials \cite{Roberts1994TheCellulose}, and a reduced Young's modulus ($E=5\times10^7$~Pa) was implemented to balance computational speed with accuracy. While a reduced Young's modulus can influence the behaviour of dense systems, values above $10^7$~Pa typically have little effect on results \cite{Paulick2015ReviewResults}. The sliding and rolling friction are the values that will be calibrated for MCC and ion exchange resin in the example use of the digital models in Sections \ref{sec:granuflow_calibration} and \ref{sec:granupack_calibration}.

    Computational resources were provided by the University of Birmingham's BlueBear HPC facility, and all simulations were conducted using Intel Xeon Platinum 8570 CPUs. Individual simulation run times varied depending on the specific instrument and particle count, ranging from approximately 14 hours to nearly two days.

    As a high-end example, simulating 240,000 ion exchange resin particles in the flow through an orifice model for 6 seconds of simulation time on 12 CPU cores took approximately 46 hours (1 day and 22 hours). In contrast, a lower-end example involved simulating 26,000 MCC particles in the tapped density analyser for 30 seconds on a single CPU core, which took 14 hours.

    \begin{table}[ht!]
    \def\arraystretch{1.5}
    \centering
    \caption{Summary of simulation parameters used in this study.}
    \label{table:sim_params}
        \begin{tabularx}{\linewidth}{ | >{\raggedright\arraybackslash}X || >{\centering\arraybackslash}X | >{\centering\arraybackslash}X | }
        \hline
        \textbf{Simulation Parameter} & \textbf{MCC} & \textbf{Ion Exchange Resin} \\
        \midrule\midrule
        Median Particle Diameter ($d_{50}$) [mm] & 1.2 & 0.58 \\
        \hline
        Material Density [kg/m$^3$] & 1460 & 960 \\
        \hline
        Timestep [s] & \multicolumn{2}{c|}{$5e^{-6}$} \\
        \hline
        Young's Modulus (E) [Pa] & \multicolumn{2}{c|}{$5 \times 10^7$} \\
        \hline
        Poisson's Ratio ($\nu$) & \multicolumn{2}{c|}{0.3} \\
        \hline
        Coefficient of Restitution ($\varepsilon$) & \multicolumn{2}{c|}{0.3} \\
        \hline
        Sliding Friction ($\mu_s$) & \multicolumn{2}{c|}{To Be Calibrated} \\
        \hline
        Rolling Friction ($\mu_r$) & \multicolumn{2}{c|}{To Be Calibrated} \\
        \hline
        \end{tabularx}
    \end{table}

\section{Digital Models Descriptions}
\label{sec:digital_models}

    \subsection{Flow Through An Orifice -- the GranuFlow}
    \label{sec:granuflow}
        
        \subsubsection{Background}
        \label{sec:real_granuflow}

        The gravity-driven flow of granular materials through an orifice captures fundamental mechanics of materials which can be useful for understanding behaviour in similar industrial processes like discharging bulk materials from a container, such as in a hopper or silo. 
        To measure the flow rate through an orifice for a granular material, lab scale instruments have been developed, the most famous example being the Hall flow meter \cite{Vlachos2011InvestigationHall-flowmeter}. 
        
\begin{figure}[!ht]
            \centering
            \includegraphics[width=0.5\linewidth]{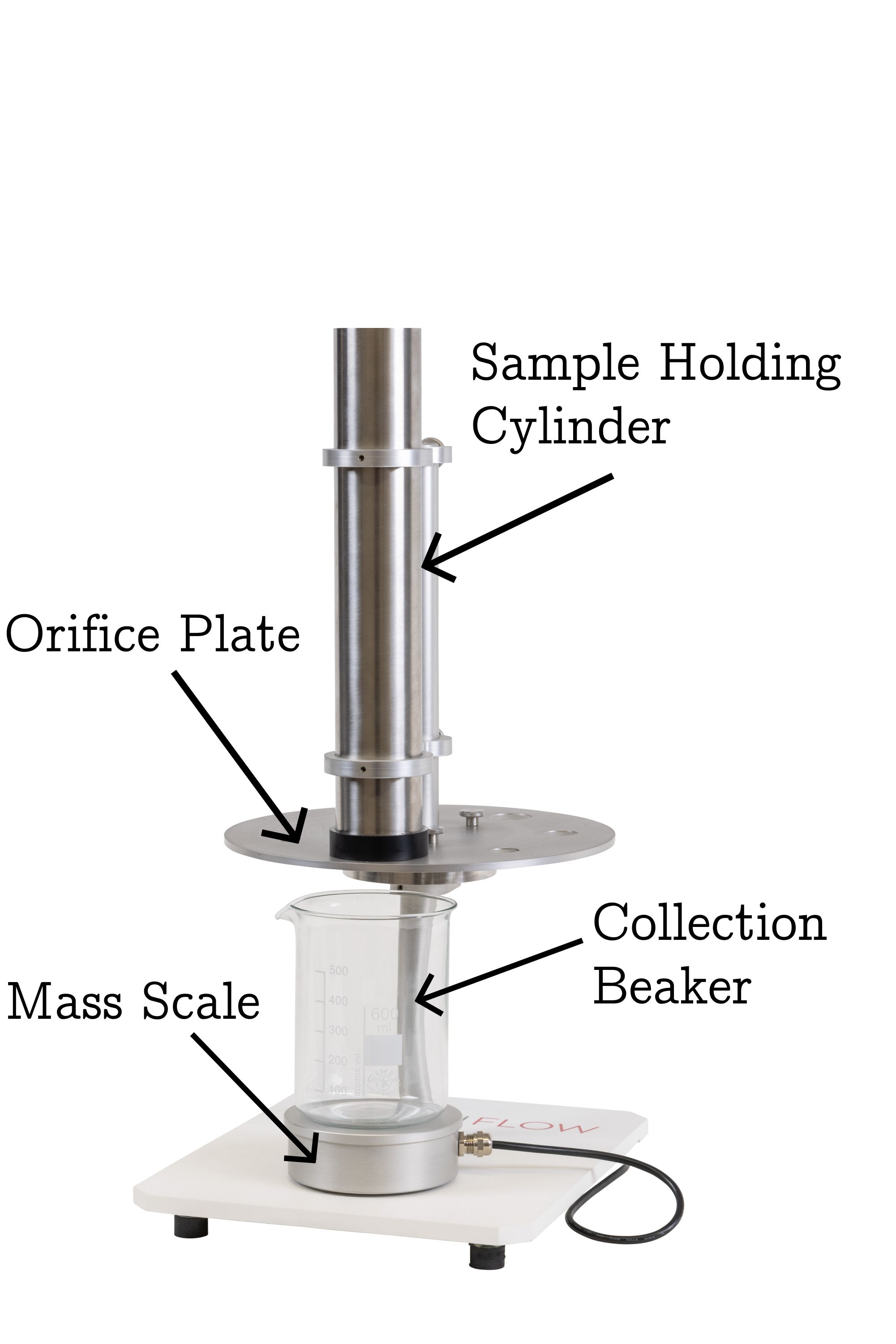}
            \caption{Image of the GranuFlow laboratory experiment.}
            \label{fig:gf_real}
        \end{figure}

        In this paper a modern example of a flow through an orifice characterisation instrument is chosen (the GranuFlow) as shown in Figure \ref{fig:gf_real}. The GranuFlow improves measurement reliability by automatically tracking the mass flow rate with an electronic scale and consists of a sample holding cylinder that holds the granular material above a set of interchangeable circular orifices with diameters of 2, 4, 8, 12, 18, 22, and 28~mm. The mass flow rate of the material being tested is measured at each orifice size. The mass flow rate can be plotted as a function of orifice size which provides a standardised way of representing data.

        This relationship between the mass flow rate and the orifice diameter can then be fitted to the empirical law proposed by Beverloo et al. \cite{Beverloo1961TheOrifices} as shown in Equation \ref{eqn:beverloo}.

        \begin{equation}
        W = c \rho_b \sqrt{g} (d_0 - kd)^{5/2}
        \label{eqn:beverloo}
        \end{equation}

        \noindent where $W$ is the mass flow rate, $\rho_b$ is the bulk density of the bulk material, $g$ is the acceleration due to gravity, $d_0$ is the orifice diameter, and $d$ is the mean particle diameter. The equation includes two dimensionless fitting parameters: $c$, a flow or discharge coefficient that is related to factors like inter-particle friction and particle shape, and $k$, a constant that accounts for the empty annulus effect. The term $(d_0 - kd)$ suggests that the effective diameter of the orifice is smaller than its geometric diameter. This is interpreted as a boundary effect where particles in an annular region of width $k \cdot d/2$ near the orifice edge are stationary or significantly slowed, not contributing to the main flow. Consequently, the term $kd$ represents the minimum orifice diameter ($D_{min}$) below which flow is stopped by stable arching.
    
        The characteristic $5/2$ power law in the Beverloo empirical equation provides a remarkably good fit to the flow rate versus orifice size curves of a wide-range of materials. Improvements have been proposed by Nedderman et al. \cite{Nedderman1992StaticsMaterials} to use the flowing density rather than bulk density of the material as this accounts for the expansion of the material as it flows out of the orifice. This was shown to reduce the range of $c$ fitting parameter significantly \cite{Nedderman1992StaticsMaterials}.

        The GranuFlow instrument, however, specifically uses a modified version of the Beverloo law though that combines the bulk density ($\rho_b$) into the $c$ term to make a new fitting term; $C$ and, the $k$ and $d$ terms are combined into $D_{min}$ which gives Equation \ref{eqn:beverloo_mod}. In this case the $C$ and $D_{min}$ are the fitting terms. This avoids needing to know the material bulk density and average particle diameter. The $C$ term from this modified equation will be used as the value for calibration of ion exchange resin in the example application of this digital model in Section \ref{sec:granuflow_calibration}. 
        
        \begin{equation}
        W = C \sqrt{g} (d_0 - D_{min})^{5/2}
        \label{eqn:beverloo_mod}
        \end{equation}

        \noindent where $W$ is the mass flow rate, $g$ is the acceleration due to gravity and $d_0$ is the orifice diameter. The equation includes two fitting parameters: $C$, a term related to particle and material properties, and $D_{min}$, a term that is related to the minimum orifice diameter that flow occurs.

        \subsubsection{Digital Model Details}

        To optimise computational efficiency, the digital model of the GranuFlow simplifies the experimental geometry. In the physical instrument, a large circular disk (``Orifice Plate'' in Figure \ref{fig:gf_real}) containing multiple orifices is rotated to select an opening. To avoid the large computational domain that this would require, this digital model replaces the disk with a linear sliding plate containing a single orifice. Each simulation contains a plate with just one orifice size but any orifice size can be used and is defined during the setup of the simulation. This design preserves the essential mechanism of an opening sliding beneath the powder column while significantly reducing the simulation domain size. All other important dimensions of the geometry were kept consistent with the physical experiment.

\begin{figure}[!ht]
        \centering
        \includegraphics[width=1\linewidth]{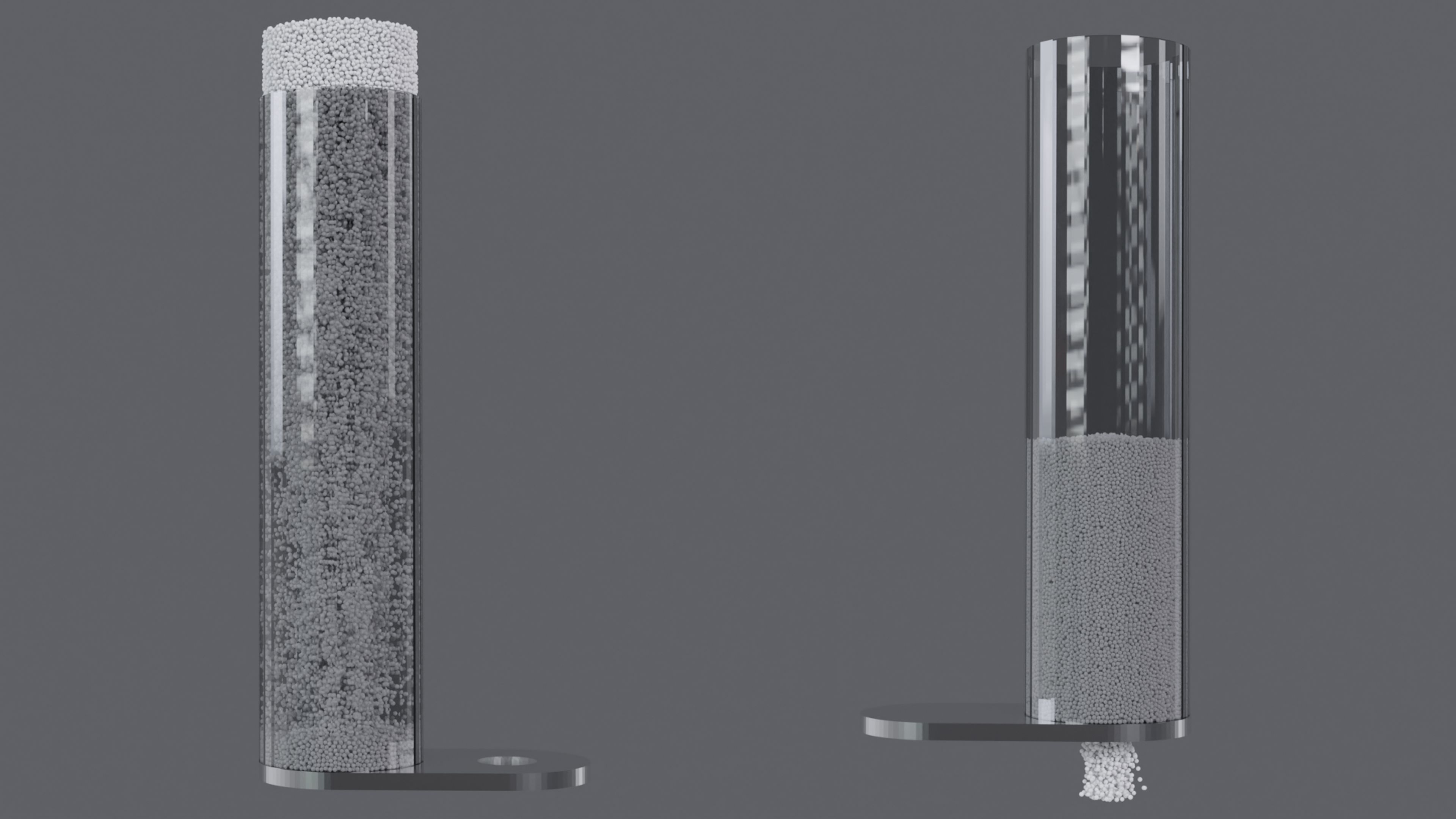}
        \caption{Key stages of the flow through an orifice digital model. Left: The silo is filled via a raining method to create a realistic packed bed. Right: The orifice is aligned, and the powder discharges.}
        \label{fig:gf_stages}
        \end{figure}

        The simulation procedure involves two main steps, as illustrated in Figure \ref{fig:gf_stages}. First, the silo is filled by generating particles in a region above the tube and allowing them to fall and settle under gravity. This raining method creates a more realistic initial packing structure than generating particles in place and it is closely representative of how the experimental instrument is filled in practice. Second, once the particles have settled, the sliding plate moves at a constant rate to position the orifice directly beneath the centre of the powder column. As the powder begins to discharge, the mass of particles remaining inside the tube is recorded at a set frequency, typically 100 Hz.
        
        In the experiment the particles are collected below the orifice on an electronic mass scale to calculate a mass flow rate. In the digital model, to enhance computational efficiency, particles are removed from the simulation integrator once they have passed a few centimetres below the orifice, as they no longer influence the flow dynamics. This means they cannot be collected below the orifice like in the experiment. Instead, the mass flow rate is calculated by recording the decrease in the mass of powder remaining within the simulation domain over time.

        This approach is directly equivalent to weighing the collected mass of powder, as in the experiment, but markedly reduces the computational cost of the simulation. By minimising the total particle count, this method allows for greater efficiency and faster run times. To align the data trend with the experiment (where collected mass increases), the remaining mass in the silo is subtracted from the initial mass to yield the total mass discharged.
        
\begin{figure}[!ht]
        \centering
        \includegraphics[width=0.98\linewidth]{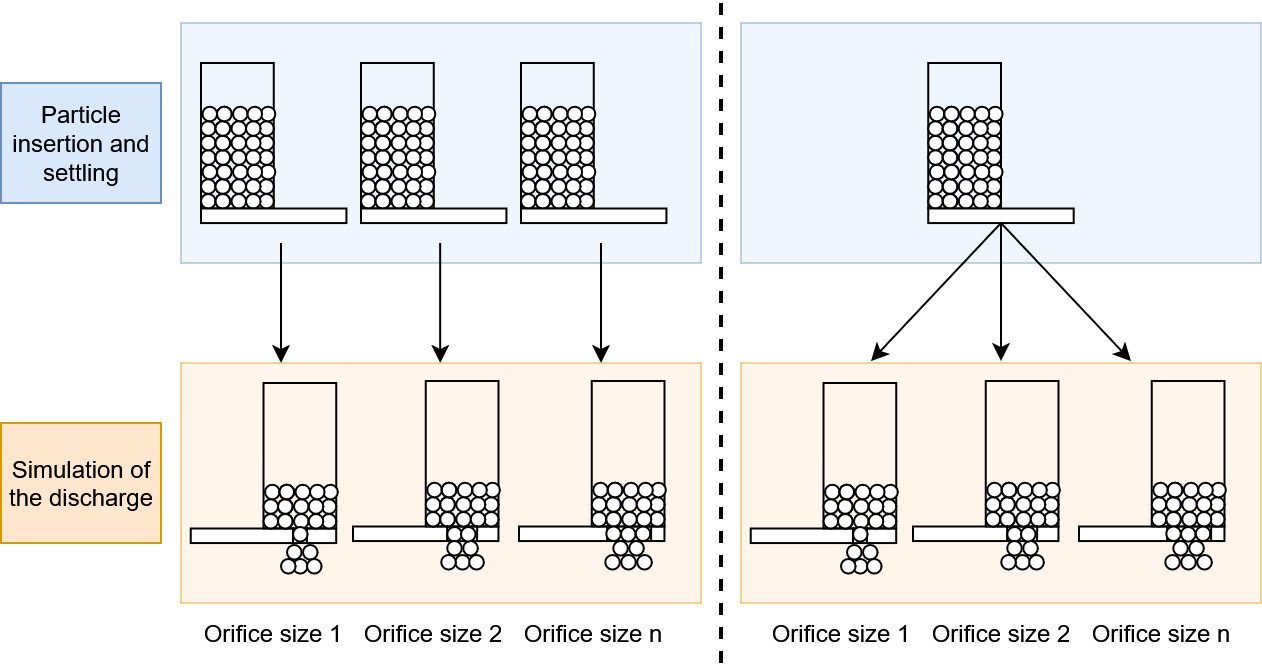}
        \caption{Diagram of the efficiency improvement achieved by running one insertion simulation for multiple orifice sizes.}
        \label{fig:gf_diagram}
        \end{figure}

        A simulation must be run for each orifice size to generate the data needed to fit the modified Beverloo equation (Equation \ref{eqn:beverloo_mod}). To avoid the inefficiency of repeating the particle insertion step for each run (left side in Figure \ref{fig:gf_diagram}), a single particle insertion and settling simulation is performed (right side in Figure \ref{fig:gf_diagram}). The state of this settled system is then saved and used as the starting point for each subsequent simulation with a different orifice size. It is important to note that this optimisation is only valid when testing a single material, as different material properties (e.g., friction, cohesion) would alter the initial packed bed. However, an advantage of this is that it ensures a fair tests between each orifice simulation as the initial conditions for each simulation for each of the different orifices sizes are completely identical.
        
        \subsubsection{Example Application}
        \label{sec:granuflow_calibration}

\begin{figure}[!ht]
        \centering
        \includegraphics[width=0.95\linewidth]{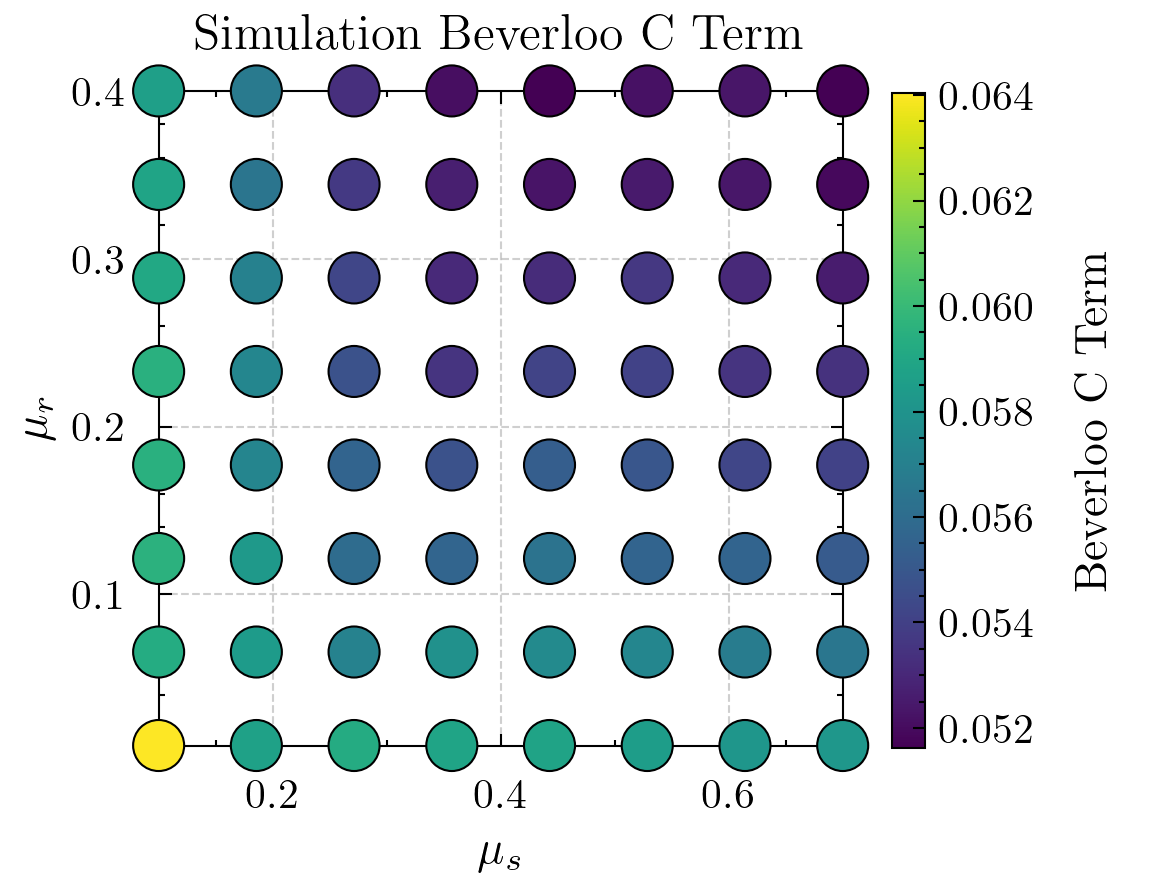}
        \caption{Fitted $C$ values for the Beverloo empirical law for each simulation.}
        \label{fig:calib_grid_c}
        \end{figure}

\begin{figure}[!ht]
        \centering
        \includegraphics[width=0.7\linewidth]{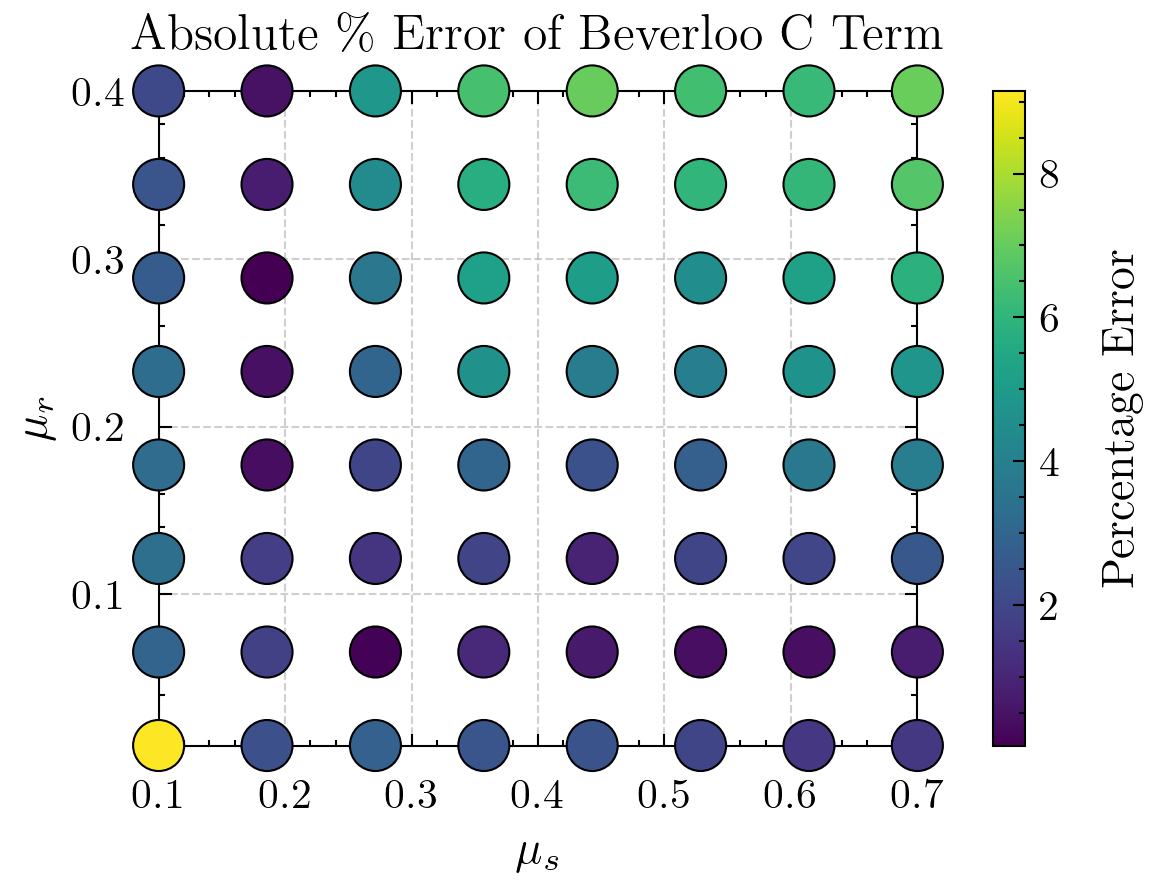}
        \caption{Deviation of fitted $C$ values between the experimentally value and simulation value for each parameter combination.}
        \label{fig:calib_grid_deviation_c}
        \end{figure}

        As an example use of the GranuFlow digital model, a simple trial-and-error calibration of the sliding and rolling friction coefficients of an ion exchange resin material in conducted in this section. To achieve this, a parameter sweep was performed, with simulations conducted for values of $\mu_s$ varying from 0.1 to 0.7 and $\mu_r$ from 0.01 to 0.4. This range was found to explore enough of the parameter space to find identify an optimum set of parameters as will be shown later in this section in Figure \ref{fig:calib_grid_deviation_c}. The Beverloo equation was fitted to the flow rate versus orifice size curve for each combination of friction values, yielding a grid of $C$ term values as shown in Figure \ref{fig:calib_grid_c}. For each combination of friction values, one initial packing simulation was run and the simulation state saved and used as the initial starting point for seven simulations for the seven different orifices used: 2, 4, 8, 12, 18, 22, 28~mm.

        Figure \ref{fig:calib_grid_c} shows that the Beverloo $C$ fitting term is dependent on the value of the frictional parameters. This is due to the flow rate of the material decreasing with increased friction strength. An increase in either the sliding or rolling friction coefficient impedes the flowability of the material, resulting in a decrease in the value of the $C$ term.

        To find the optimal parameters, the relative percentage error ($\Delta = 100 \cdot \frac{|C_{exp}-C_{sim}|}{C_{exp}}$) is calculated between the value measured in the simulation for each set of $\mu_s$ and $\mu_r$ and the experimental ground truth, which was found to be $C=0.0586$ (Table \ref{table:experimental_measurements}). The resulting relative percentage error values are plotted against the parameter space in Figure \ref{fig:calib_grid_deviation_c}. By identifying the minimum error, the most suitable combination of sliding and rolling friction can be chosen to be; $\mu_s=0.53$, $\mu_r=0.066$. For these parameters, the value of the fitted $C$ term agrees with the experimental with a relative error of 1.1\%. Identifying, the optimal parameter set was tricky though due to multiple different parameter sets producing a small error.
        
         More advanced calibration method that make use of machine learning and optimisation techniques \cite{Nicusan2022ACCES:Simulation, Windows-Yule2022CalibrationSystems} or using multiple instruments together would most likely help to constrain the solution and identify one unique ``true'' parameter set. However, this simple calibration is only meant to show an example application of the digital model and advanced calibration methods are not the focus of this paper.

    \subsection{Tapped Density Analyser -- The GranuPack}
    \label{sec:granupack}

        \subsubsection{Background}
        \label{sec:tapped_density}
        
\begin{figure}[!ht]
            \centering
            \includegraphics[width=0.5\linewidth]{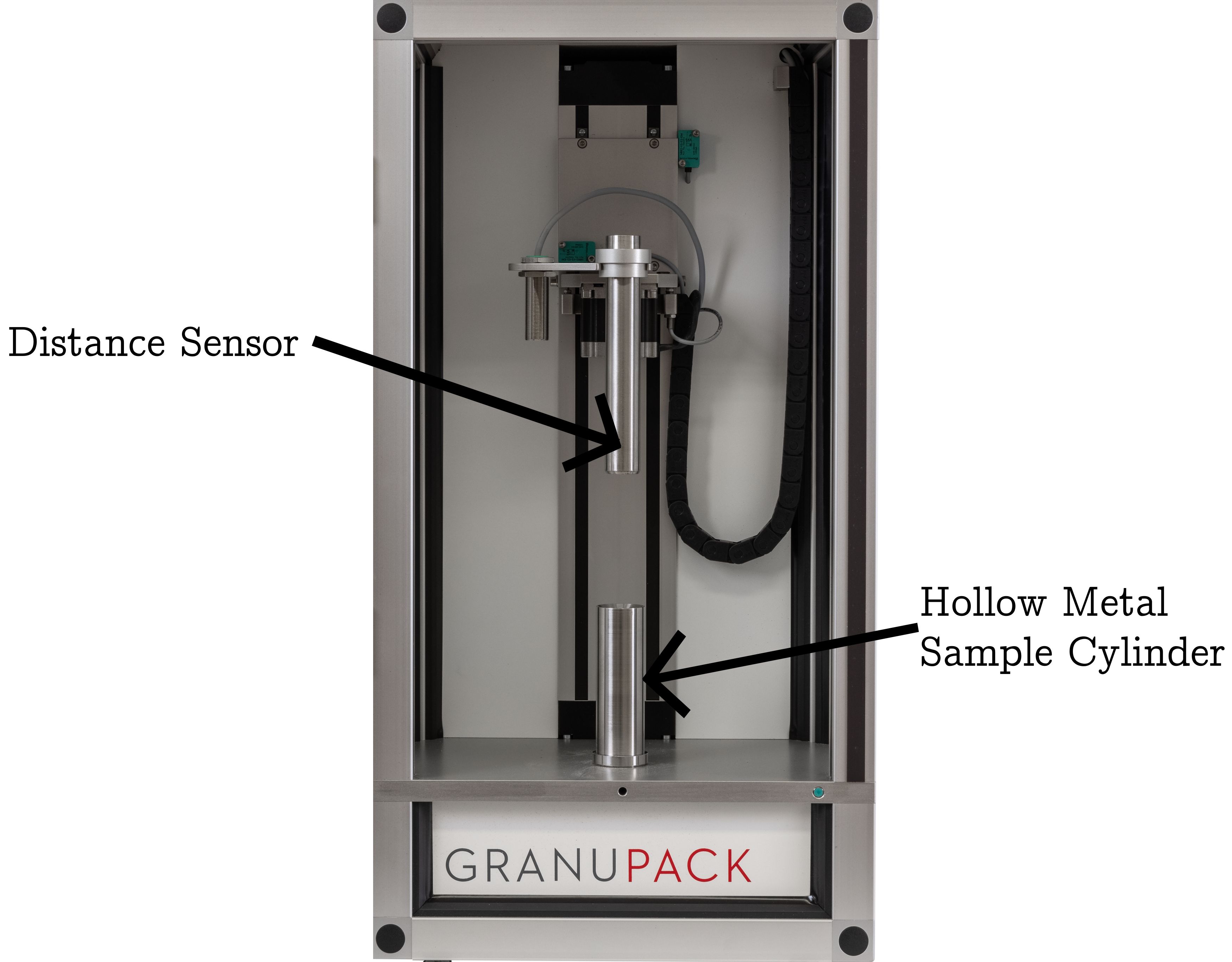}
            \caption{Image of the GranuPack laboratory experiment.}
            \label{fig:gp_real}
        \end{figure}

        While the GranuFlow digital model provides details of the behaviour of a bulk material under relatively dynamic particle rearrangement, the GranuPack, in Figure \ref{fig:gp_real}, explores how powders rearrange and consolidate under controlled energy input in a quasi-static case. The packing behaviour and compressibility of a granular material provide useful insights into handling characteristics during storage, transport, and processing under vibration \cite{Nowak1998DensityMaterials}.

        Powder compaction can be assessed by using several metrics (that will be discussed below) that compare the initial bulk density ($\rho_{bulk}$) of a powder to its final tapped density ($\rho_{tapped}$) \cite{ASTM2022B527-22:Compounds}. The bulk density is the density of the powder in a loose, unsettled state, whereas the tapped density is the density achieved after the powder has been consolidated under a standardised energy input. The extent to which a powder compacts is dependent upon its properties, predominantly inter-particle friction \cite{Yu1994PackingTapping} but also other properties such as particle shape, size distribution and cohesion.
    
        
        The GranuPack instrument consists of a hollow metal sample cylinder which holds the material being tested. The tapping mechanism is driven by a rotating snail cam; as the cam turns, it slowly raises the cylinder and then allows it to fall abruptly under gravity, creating a single, controlled tap. The cam's rotation speed is controlled to produce a tapping frequency of 1~Hz (i.e. one tap every second), and the drop height can be set to either 1~mm or 3~mm. 
        
        To ensure reproducible and accurate results, the GranuPack instrument uses a laser distance sensor to precisely measure the height of the powder column during a controlled tapping sequence. To start a measurement run, the metal cylinder is first loaded with 35~ml of material via a funnel and to ensure an accurate height reading from the laser distance sensor, a small, lightweight, diabolo-shaped piece is carefully placed on top of the material. After this, the tap sequence is started. 

        This automated approach provides an advantage in data resolution. While traditional methods, which rely on manual observation, might only record the height at a few intervals (e.g., after every 100 taps) as they require a human to make observations and record them, this electronic logging can record the packing state after every single tap. This allows for the capture of the entire packing curve, offering a deeper insight into the powder's consolidation dynamics. This is also useful if a user wishes to compare the packing curves directly to the digital model (discussed in the next section) as both can save data at every tap.
    
        The relationship between the initial bulk density and final tapped density is commonly expressed using the Hausner ratio ($H_r$) and Carr's compressibility index ($C_I$), as proposed by Hausner \cite{Hausner1967FrictionPowder} and Carr \cite{Carr1965EvaluatingSolids}, respectively. These indices are calculated as follows:

        \begin{equation}
        H_r = \frac{\rho_{tapped}}{\rho_{bulk}}
        \label{eqn:hausner}
        \end{equation}

        \begin{equation}
        C_I (\%) = 100 \times \frac{\rho_{tapped} - \rho_{bulk}}{\rho_{tapped}}
        \label{eqn:carr}
        \end{equation}

        \noindent where $\rho_{tapped}$ is the tapped bulk density and $\rho_{bulk}$ is the initial untapped bulk density.
    
        For the Hausner ratio, a value close to 1 typically indicates good flowability, while values significantly greater than 1 (typically > 1.25), after 500 taps, suggests a higher cohesion and poor flowability. For the Carr's Index, a value less than 18\%, after 500 taps, is characteristic of free-flowing powders, whereas a high value > 22\%, after 500 taps, is associated with cohesive, poorly flowing materials \cite{Traina2013FlowMeasurement}.
        
        \subsubsection{Digital Model Details}
        \label{sec:granupack_digital_model}

\begin{figure}[!ht]
        \centering
        \subfloat[Full GranuPack digital model.]{\includegraphics[width=0.48\textwidth]{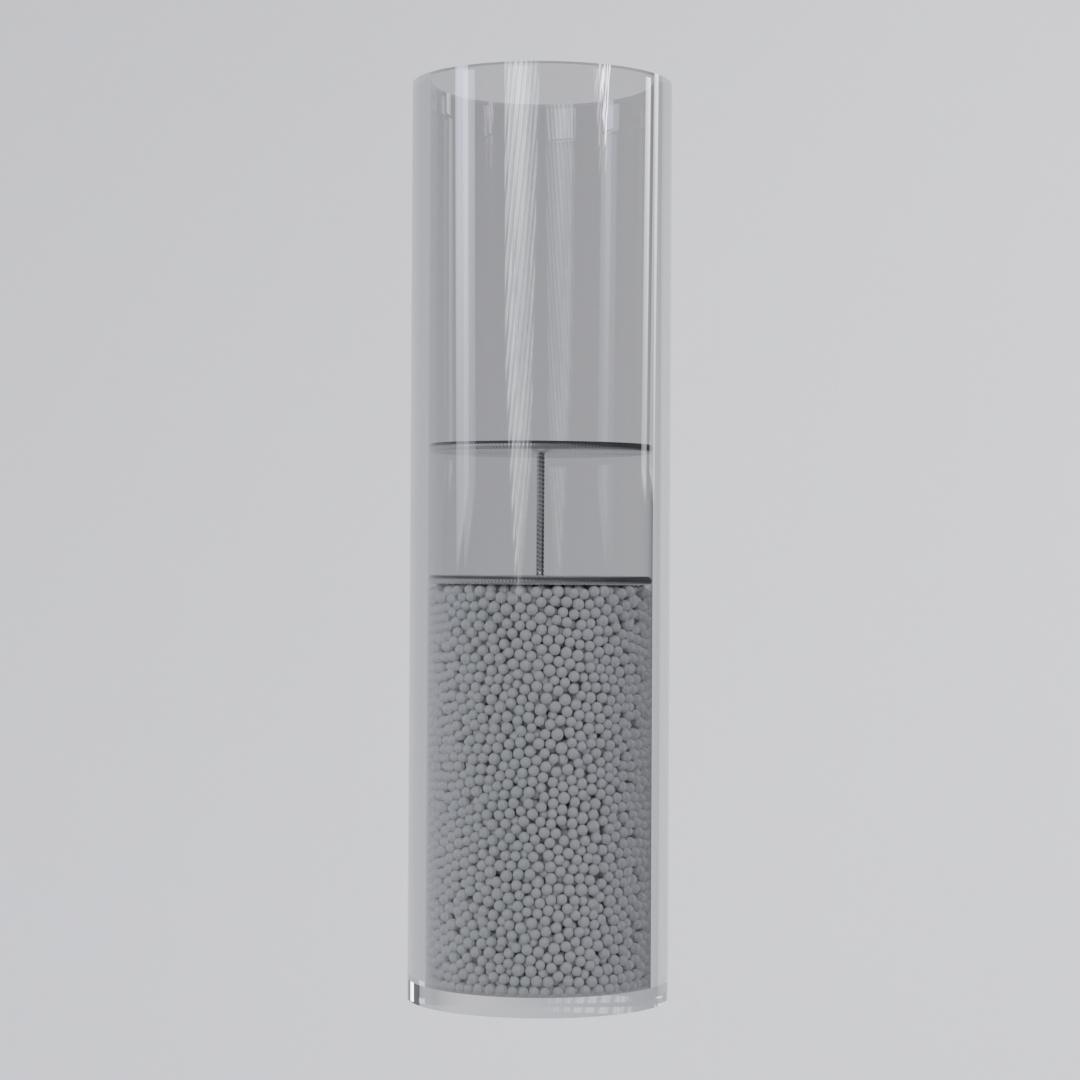}\label{fig:pack_full_digital_model}}
        \hfill
        \subfloat[Multisphere diabolo-shaped piece.]{\includegraphics[width=0.48\textwidth]{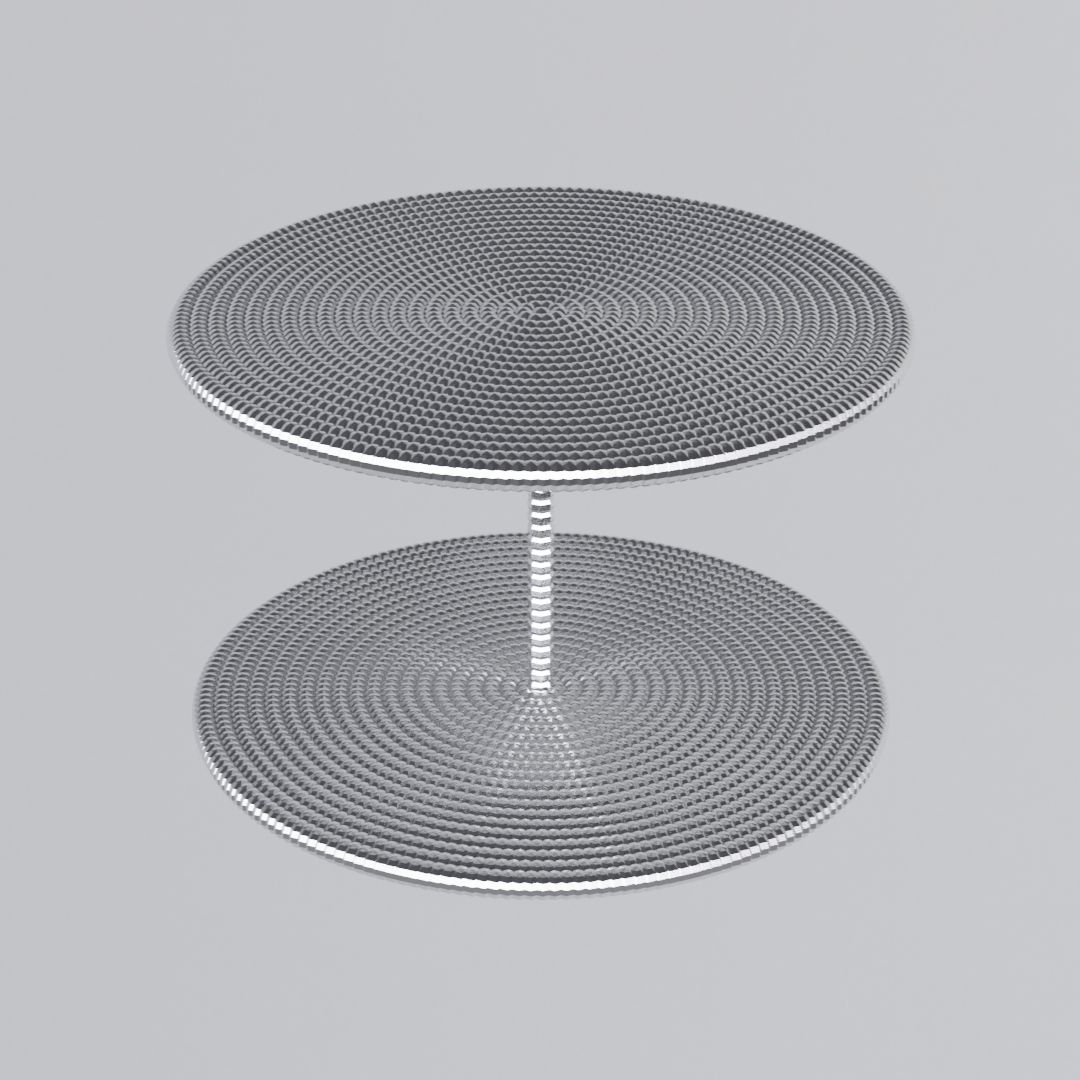}\label{fig:pack_diabolo_model}}
        
        \caption{Digital model of the GranuPack.}
        \label{fig:pack_digital_model}
        \end{figure}

        The digital model of the GranuPack tapped density analyser replicates the exact geometry of the physical system, including the metal diabolo-shaped piece that sits on top of the material bed (Figure \ref{fig:pack_digital_model}). This piece was constructed as a single rigid body using the multisphere method, with the same mass and dimensions as its real-world counterpart.

        In the simulation, the diabolo's primary role is to accurately replicate the physical system and any effects the piece may have on the packing behaviour of the bulk material. Unlike in the physical experiment, it is not used for measuring the bed height, as exact particle positions can be retrieved directly from the DEM data.


        The simulation begins by raining a predetermined number of particles into the cylinder to match the initial experimental volume of 35~ml. This method creates a realistic initial packing structure, as it imitates the material insertion process in the experiment. Once the kinetic energy of the particles has stabilised, the multisphere diabolo piece is inserted just above the highest particle in the bed, leaving a small safety gap. This prevents additional packing that could be caused by the piece impacting the material and imitates the experiment where the diabolo piece must be gently placed on top of the material to also prevent unwanted packing.

        The tapping sequence is then initiated. To improve computational efficiency, the process is simplified compared to the experiment. In the model, the cylinder geometry falls under gravity to create a tap, but it remains at its new, lower position instead of moving back up. A new tap is initiated as soon as the kinetic energy of the particle system has stabilised. This allows for a much higher tapping frequency (typically 5–10~Hz) than in the physical experiment, greatly shortening the required simulation time. The tap frequency can be set to 1~Hz, though, to conduct tests to ensure the increased tap frequency does not change the results. To further optimise performance, the total downward travel of the cylinder is pre-calculated, allowing the simulation domain to be minimised (e.g. if the drop height per tap is 1~mm and 150 taps are conducted then the total downward travel can be calculated as $150 \cdot 1=150$mm and the simulation domain size set accordingly). The drop height, per tap, can be set to any size but as a default either 1 or 3~mm is set, matching the experimental options.

        After each tap, the bulk density of the material is calculated. The total mass is found by summing the volume of all individual particles and multiplying by the material density. The bulk volume is calculated from the height of the powder bed within the cylinder. The bed height is determined from the vertical positions of the highest and lowest particles.

        Once the prescribed number of taps is complete, the Hausner ratio and Carr's compressibility index are calculated from the initial and final bulk densities using the same equations (\ref{eqn:hausner} and \ref{eqn:carr}) as in the physical experiment.

        \subsubsection{Example Application}
        \label{sec:granupack_calibration}

        Following the same approach as the GranuFlow example application in Section \ref{sec:granuflow_calibration}, the digital GranuPack model was used to calibrate the sliding ($\mu_s$) and rolling friction ($\mu_r$) coefficients for the MCC material described in Section \ref{sec:experimental_materials}. A total of 64 simulations were run, corresponding to an 8$\times$8 grid of parameter values. The $\mu_s$ values ranged from 0.1 to 1.0, and the $\mu_r$ values ranged from 0.01 to 0.5. Each simulation was run for 150 taps, matching the experimental procedure, as the bulk density showed negligible change beyond this point.

\begin{figure}[!ht]
        \centering
        \includegraphics[width=0.7\linewidth]{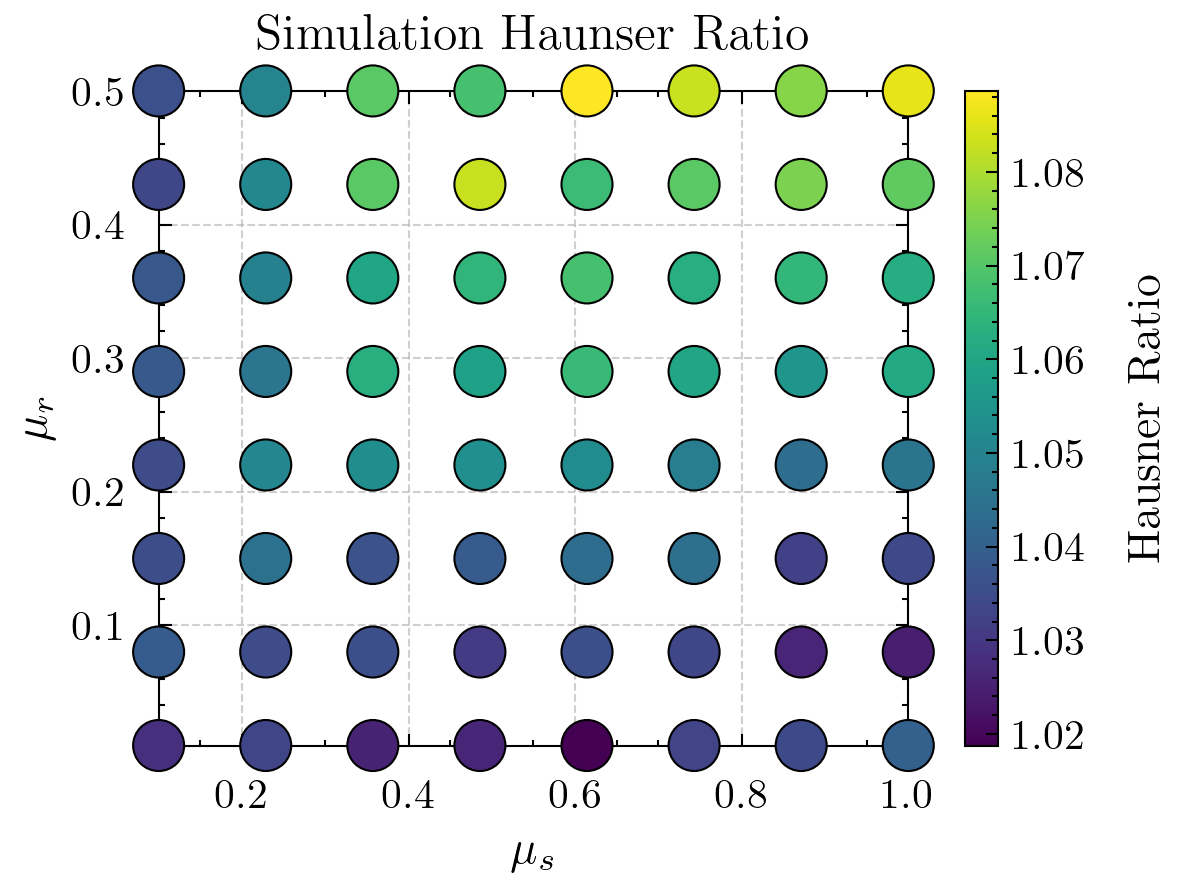}
        \caption{Measured values of Hausner ratio in the simulations for each set of friction value combinations.}
        \label{fig:calib_grid_hr}
        \end{figure}

        The Hausner ratio was calculated for each parameter combination and is plotted in Figure \ref{fig:calib_grid_hr}. As shown, there is a strong positive correlation between the Hausner ratio and both the sliding and rolling friction coefficients. This is the expected trend, as higher inter-particle friction leads to a lower initial particle bulk density, leading to a larger ratio between the initial bulk density and the final tapped density and thus higher Hausner ratio.

\begin{figure}[!ht]
        \centering
        \includegraphics[width=0.7\linewidth]{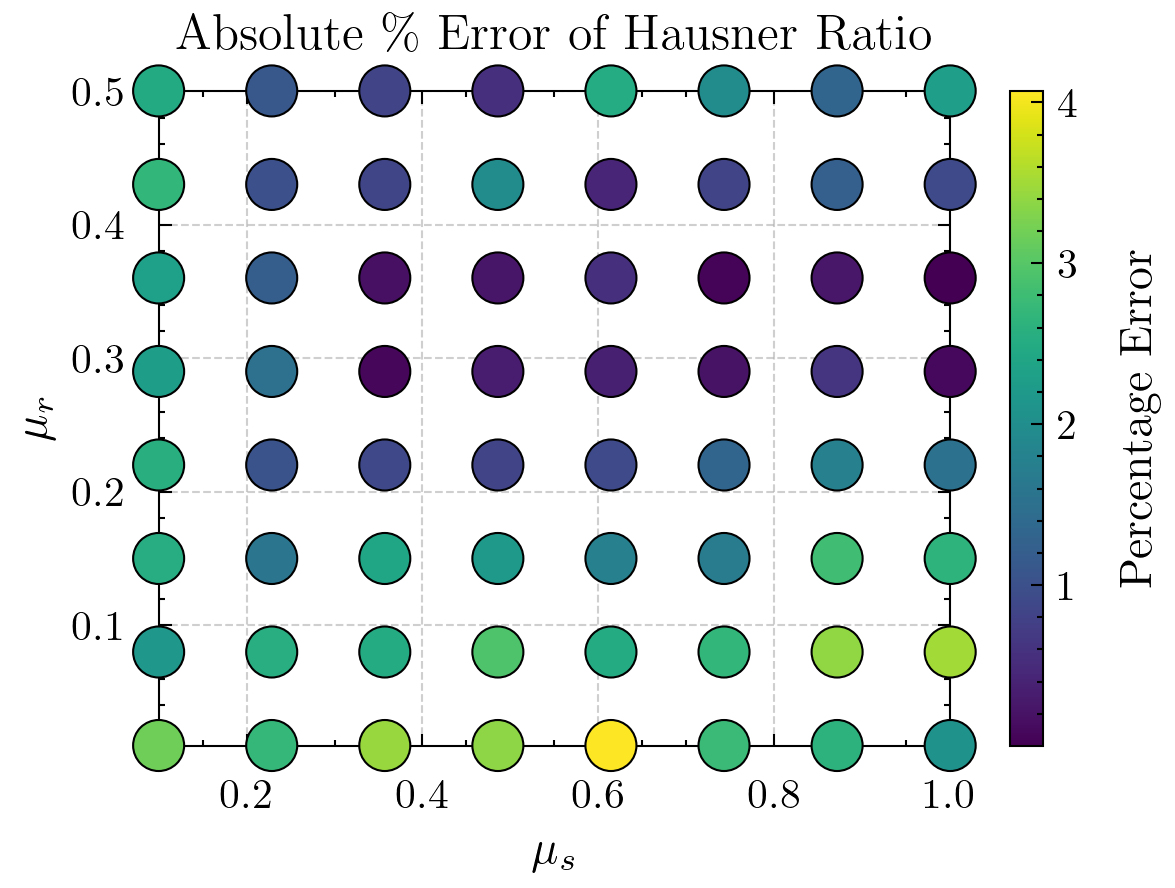}
        \caption{The absolute percentage error of Hausner ratio in the simulations to the experiment.}
        \label{fig:calib_grid_deviation_hr}
        \end{figure}

        Figure \ref{fig:calib_grid_deviation_hr} shows the absolute percentage error in the Hausner ratio between the experimental measurement and measurements in each of the 64 simulations. Here a lower value is better, indicating the error between Hausner ratio in the simulation and experiment is low. The optimal parameter combination was found to be $\mu_s=0.74$, $\mu_r=0.29$ although this was hard to identify in this simple, single instrument calibration approach due to many combinations giving a low error. More advanced calibration methods that use multiple instruments together would likely better constrain the result, but is beyond the scope of this work as the primary focus is the digital models themselves. For these parameters, the value of the Hausner ratio agrees with the experiment with a relative error of 0.3\%.

\section{Cross Validation}
\label{sec:cross_validation}

To assess the robustness and transferability of the calibrated parameters, a cross-validation was performed between the two digital models. This tests whether parameters calibrated using the GranuFlow digital model can accurately predict bulk behaviour in the GranuPack digital model and vice versa.

First, the friction coefficients for the ion exchange resin, calibrated using the GranuFlow digital model in Section \ref{sec:granuflow_calibration} ($\mu_s = 0.53$ and $ \mu_r = 0.066$), were used as inputs for the GranuPack simulation. The resulting simulated Hausner ratio was then compared to the experimental data. The simulation produced a Hausner ratio of $H_r=1.056$, which agrees with the experimental measurement of $H_r=1.052$ with an relative percentage error of 0.57\%.

Similarly, the parameters for MCC, calibrated using the GranuPack model in Section \ref{sec:granupack_calibration} ($\mu_s = 0.74$ and $\mu_r = 0.29$), were validated in the GranuFlow digital model. The simulation had a Beverloo $C$ fitting term of 0.0591 which had a 4.17\% relative percentage error to the $C$ value fitted to the experimental data.

Table \ref{table:validation_summary} provides a summary of the validation cases.

\begin{table}[h!]
    \centering
    \caption{Summary of cross-validation cases comparing experimental measurements with predictions from the digital models using transferred parameters.}
    \label{table:validation_summary}
    \begin{tabular}{|| p{3.0cm} | p{2.5cm} | c | c ||}
        \hline
        \textbf{Validation Measurement} & \textbf{Material} & \textbf{Experiment} & \textbf{Digital Model} \\ [0.2ex] 
        \hline\hline
        Hausner Ratio & Ion Exchange Resin & 1.052 & 1.056 \\ 
        \hline
        Beverloo C Fitting Term & MCC & 0.0575 & 0.0599 \\ 
        \hline
    \end{tabular}
\end{table}

\section{Summary and Conclusions}

In this work, a set of digital models have been developed and validated for two widely-used powder characterisation instruments: the GranuFlow (flow through an orifice tester) and the GranuPack (tapped density analyser). These models expand the open-source library developed in previous papers in this series, providing access for researchers and engineers to a wide variety of well documented and free to use simulations. Both the operation of the models and the processing of their results have been rigorously replicated to match the experiments. Across the three papers in this series, a comprehensive suite of digital models of powder characterisation instruments has been developed. Together, these models explore a wide spectrum of granular behaviour, including consolidated shear behaviour, unconfined dynamic flow, static stability and now particle rearrangement, compaction and gravity driven flow. 

Providing a standardised suite of digital models to the wider DEM community offers a set of peer-reviewed simulations that will improve the quality and repeatability of calibration studies between different research groups. To support adoption and further innovation, all models and auxiliary codes are openly available at the University of Birmingham Positron Imaging Centre GitHub repository (\href{https://github.com/uob-positron-imaging-centre}{\color{black} https://github.com/uob-positron-imaging-centre}) and the Granutools Gitlab (\href{https://gitlab.com/gdigitaltwins}{https://gitlab.com/gdigitaltwins}).


To demonstrate the application of the digital models, a simple DEM calibration was performed for both MCC and ion exchange resin to determine sliding and rolling friction coefficients for the two.

Furthermore, to test the transferability of the parameters, a cross-validation was performed: the values obtained from one digital model were used to predict the results in the other, which were then compared against the corresponding experimental data. For example, the MCC parameters, calibrated using the tapped density analyser, were tested in the flow through an orifice model.

This cross-validation showed reasonable results. The MCC validation yielded a relative percentage error of 4.71\% for the fitted Beverloo $C$ term. For the ion exchange resin, calibrated in the flow through an orifice tester, the validation in the tapped density model yielded a relative percentage error of 0.57\% for the Hausner ratio.

\section*{Acknowledgements}

Authors acknowledge financial support received from the Centre for Doctoral Training in Formulation Engineering (EPSRC grant number EP/S023070/1) and Granutools. Computational resources have been provided by the University of Birmingham BlueBear facility (see http://www.birmingham.ac.uk/bear for more details) and the Sulis Tier 2 HPC platform hosted by the Scientific Computing Research Technology Platform at the University of Warwick. Sulis is funded by EPSRC Grant EP/T022108/1 and the HPC Midlands+ consortium.

\bibliography{references}

\end{document}